\documentclass[12pt]{article}
\usepackage{amsfonts,amssymb,amscd}
\usepackage[english]{babel}
\newcommand{\be}{\begin{equation}\label}
\newcommand{\ee}{\end{equation}}

\newcommand{\wh}{\widehat}
\def\B{{\mathbb B}}
\def\C{{\mathbb C}}
\def\Q{{\mathbb Q}}
\def\R{{\mathbb R}}
\def \GC {{\it Gravit. \& Cosm.}}
\begin{document}

\begin{center}

{\bf ALGEBRODYNAMICS IN COMPLEX SPACE-TIME AND \\
	THE COMPLEX-QUATERNIONIC ORIGIN OF MINKOWSKI GEOMETRY}

\bigskip

{\bf Vladimir V. Kassandrov}

\end{center}

\bigskip

\noindent
{\footnotesize {\bf Abstract.}
We present a scheme of biquaternionic algebrodymamics based on a nonlinear
generalization of the Cauchy-Riemann ``holomorphy'' conditions considered
therein as fundamental field equations. The automorphism group $SO(3,\C )$
of the biquaternion algebra acts as a proper Lorentz group on a real space
whose coordinates are bilinear in the complex coordinates of biquaternionic
vector space. A new invariant of Lorentz transformations then arises --- the
geometric phase. This invariant can be responsible for the quantum
properties of particles associated in this approach with field
singularities. Some new notions are introduced, related to ``hidden''
complex dynamics: ``observable'' space-time, the ensemble of identical
correlated particles-singularities (``duplicons'') and others.}

\section{Introduction. Geometric structures and fields \\
in complex space-time}

The complexification $\C \bf M$ of Minkowski space-time $\bf M$ arises
permanently in the framework of general relativity and field theories.
The generally accepted complex nature of physical fields also implies the
most natural realization in complex coordinate space. The holomorphic
structure of field and coordinate spaces can essentially simplify the
calculations of string diagrams and, after reduction to a Euclidean sector,
ensures convergence of functional integrals.

It is well known that the ``virtual'' dynamics in  $\C \bf M$ gives rise to
a peculiar real dynamics in its Minkowski ``cut''. For example, null
congruences with {\it twist\/}, the Kerr congruence among them, may be
viewed as generated by a point ``charge'' located at some point of $\C \bf
M$, and the Kerr parameter $a$ has the meaning of charge separation from the
real slice $\bf M$.

More generally, E.T. Newman et al. \cite{Newman,Lind,Newman2} have
demonstrated that the {\it complex null cone} of a point charge moving along
an arbitrary world line in $\C \bf M$ forms a {\it shear-free null
congruence} of rays which is related to asymptotically flat solutions of
Maxwell, Einstein or Einstein-Maxwell equations with interesting
particle-like properties, in particular, with the gyromagnetic ratio equal to
that of a Dirac fermion (recall that Carter \cite {Carter} was the first to
notice this property of Kerr's singular ring).

Note that it is just the {\it shear-free} null congruences that have
defining equations {\it holomorphic} with respect to the principal null
2-spinor field $\lambda_A$ of the congruence (i.e. do not contain the
complex-conjugated spinor $\lambda_{A'}$):
\be{SFC}
	\lambda^{A}\lambda^{B}\nabla_{AA'} \lambda_B =0, ~~ A,B,... =0,1,
\ee
and can therefore be naturally extended analytically to the whole $\C \bf M$
space. A set of interesting physical fields, in particular, the Maxwell one,
has been associated with such congruences by I. Robinson \cite{Rob}, P. Tod
\cite{Tod}, E.T. Newman \cite{Newman3} and others. The electromagnetic field
(as well as the complex Yang-Mills field), with nontrivial gauge symmetries
and with a {\it self-quantized electric charge\/} of its bounded
singularities --- sources, has been introduced in our works
\cite{AD,GR95,Vestnik,Jos,Sing} in the general framework of the so-called
{\it algebrodynamical} paradigm.

In {\it algebrodynamics} (see, e.g., \cite{AD,GR95,Acta,Pavlov,GRG} and
references therein) one seeks a {\it numerical\/}, abstract origin of the
physical laws and space-time geometry, i.e., a sort of {\it the Code of
Nature\/}, reproducing, at the most elementary level, the {\it genetic\/}
code. In particular, the exceptional algebra of {\it quaternions\/} $\Q$ can
be accepted as the fundamental {\it space-time algebra}.  Since its
invention by Sir William Hamilton in 1843, it is known that $\Q$ possesses
a perfect structure for describing our 3D space geometry since its {\it
automorphism\/} group is (2:1) the group of 3-rotations $SO(3)$.

Moreover, Hamilton was in fact the first, nearly 50 years before Minkowski,
to introduce the concept of unique {\it space-time\/} via identifying the
fourth ``real'' unit of $\Q$-algebra with physical time. It was quite
natural in the times of Hamilton since this unit remains invariant under
the $SO(3)$ transformations as it should be for Newtonian absolute time.
However, the relative nature of physical time and the fundamental Lorentz
invariance established later in special relativity (SR) have come to an
evident contradiction with the structure of the $\Q$ algebra.

To escape this difficulty, the authors usually deal with a complex extension
of $\Q$, the algebra of {\it biquaternions\/} $\B$. Then, on a coordinate
subspace of $\C^4$ vector space of $\B$, one reveals the metric of real
Minkowski space $\bf M$, and the theory on this slice is completely
Lorentz-invariant.

In our previous works, we also used this scheme and thus constructed a
Lorentz-invariant algebraic field theory. The fundamental physical field is
therein represented by ``hyper-holomorphic'' {\it biquaternionic}-valued
functions which satisfy the primary field equations, {\it the generalized
Cauchy-Riemann conditions} proposed in 1980 \cite{VINITI} (see also
\cite{Acta,Tallinn}).  Remarkably, the latter are {\it nonlinear\/} owing to
the property of {\it noncommutativity\/} of $\B$ and are thus capable of
describing (self-) interacting fields. For details of $\B$-algebrodynamics
we refer the reader to our papers \cite{AD,GR95,Protvino,Jos,Eik}.

However, the restriction of four complex coordinates of $\B$ to a real
Minkowski subspace (which does not even form a subalgebra in $\B$) seems
rather artificial. Moreover, the problem is even more difficult: we do not
know any algebra whose automorphism group would be isomorphic to the Lorentz
group of SR (or naturally contain the latter as its subalgebra). It is
clear that without solving this problem one would not be able to propose a
consistent realization of the algebraic ({\it algebrodynamical}) programme.

Meanwhile, there is a peculiar property of $\B$ algebra that distinguishes
it as a probable algebra of physical space-time: {\it its automorphism group
$SO(3,\C)$ is 6-parametric and isomorphic to the (proper) Lorentz group\/}.
However, it acts on the 3D complex (6D real) space which, at first glance,
has no relation to 4D Minkowski space. Nonetheless, the isomorphism\\ $SO(3,1)
\simeq SO(3,\C )$ has been used in a number of works, e.g., in the {\it
quaternionic theory of relativity\/} developed by A.P. Yefremov \cite
{Efremov1,Efremov2}. In this approach, one deals with three spatial and
three temporal coordinates represented by the real and imaginary parts of
(the linear space of) the complex vector units $I,J,K$ of $\B$, respectively,
whereas the fourth scalar unit is ``frozen'' and is not involved in the
physical dynamics. To reduce the 3D time to physical one-dimensional time,
special orthogonality conditions are imposed.

Alternatively, in our recent papers \cite{Levich,PIRT}, the geometry of
a {\it complex null cone\/} (CNC) in the $\C^4$ vector space of $\B$ has
been exploited.  Points separated by a null complex interval have equal {\it
twistor fields\/} and, as a consequence, {\it their dynamics is correlated}.
In turn, the 6D ``observable'' space of the CNC naturally decomposes into 4D
physical space-time and an orthogonal space of a 2-sphere (internal {\it
spin\/} space). A physical time interval is defined by this as the {\it
whole distance passed by a point particle in the complex space $\{Z_\mu\}\in
\C $ with respect to the {\it Euclidean} metric $\Sigma |Z_\mu|^2$}.

However, all these and other perspective schemes (among which, an
interesting treatment of (3+3) geometry by I.A. Urusowski \cite{Urus}
should be certainly mentioned) suffer from an evident lack of coordination
with SR since the role of Lorentz symmetries is vague. In this paper
(section 2), we construct a nonlinear mapping of the $\C^4$ vector space of
$\B$ algebra into the {\it timelike\/} (including its null cone boundary)
subspace of real Minkowski space on which the $\B$-automorphism group
$SO(3,\C )$ acts exactly as the Lorentz group. In other words, we show that
physical (3+1) space-time, together with its internal causal structure,
is in fact encoded in the internal properties of the biquaternion algebra
which could thus be really treated as the space-time algebra.

In section 3 we briefly consider algebrodynamics in complex quaternionic space
and its ``projection'' onto Minkowski space-time based on the above
constructed mapping. The existence of a ``hidden''  complex space leads to a
number of new concepts related to the dynamics of particles-singularities
and, probably, to their quantum properties. In particular, we briefly
discuss the structure of ``observable'' space-time, the procedure of
formation of an ensemble of identical correlated ``duplicons'' and the
``evolutionary'' meaning of the proper complex time coordinate.

\section{Biquaternion algebra as 
the origin of Minkowski space-time}

The complex quaternion (biquaternion) algebra $\B$ has an exact
representation as the full $2\times 2$ complex matrix algebra, i.e., the
following one ($\forall Z \in \B$):
\be{represent}
	Z = \left(\begin{array}{cc}
		u & w \\ p & v
				\end{array}\right)
	= \left(\begin{array}{cc}
		z_0+z_3 & z_1 - iz_2\\  z_1 + iz_2 & z_0 - z_3
				\end{array} \right).
\ee
Here $\{u,w,p,v\} \in \C $ or $\{z_0, z_1, z_2, z_3\}\in \C $
are four {\it complex} coordinates, matrix or Cartezian, respectively.

Under the $3\C $-parameter transformations of the (internal)
automorphism group $SO(3,\C)$, preserving multiplication in $\B$, one has
\be{aut}
	Z \mapsto {\overline Z} = M * Z * M^{-1},
\ee
$\forall M \in \B: \det M = 1$. Three ``vector'' coordinates $\{z_a\},\
a=1,2,3$ undergo 3D complex rotations whereas the ``scalar'' coordinate
$z_0$ corresponding to the trace part of the $Z$-matrix remains invariant.
Another invariant of these transformations is the determinant of
the $Z$-matrix, the complex (indefinite) metric in $\C ^4$,
\be{deter}
	\gamma = \det Z = (z_0)^2 - (z_1)^2 - (z_2)^2 - (z_3)^2 = {\rm inv}.
\ee
Let us here stress that, despite the representation used above, this
complex metric has no explicit relation to the Minkowski metric.

Due to invariance of the trace coordinate $z_0$, one can take as the
second independent invariant the 3D complex metric
\be{3Dmetrics}
	\sigma = (z_1)^2 + (z_2)^2 + (z_3)^2 ,
\ee
which represents {\it two real-valued invariants}. The main of them
is the (square of the) absolute value $S\in \R_+$:
\be{Minvar}
S^2 = \sigma \sigma^* = (\Re\sigma)^2+(\Im\sigma)^2 = 
(\vert {\bf p} \vert^2 -\vert {\bf q} \vert^2)^2 + (2 \vert {\bf p} 
\cdot {\bf q}\vert)^2,
\ee
where the two 3-vectors ${\bf p}$ and ${\bf q}$ represent the real and
imaginary parts, respectively, of a complex coordinate 3-vector 
\be{3vector}
	z_a = p_a + i q_a,~~ a=1,2,3;
\ee
the notations $|...|$, $\cdot$ and $\times$ (see below) stand for the usual
operations of 3-vector algebra: the modulus, scalar and vector products,
respectively.

By simple transformations using vector algebra, the expression
(\ref{Minvar}) can be brought to the following equivalent form:
\be{MMinvar}
	S^2 = (| {\bf p} |^2 + | {\bf q} |^2)^2 -
	(2 | {\bf p} \times {\bf q}|)^2 \equiv T^2 - | {\bf X} |^2,
\ee
where the positive-definite coordinate
\be{time}
	T = | {\bf p} |^2 + | {\bf q} |^2
\ee
can be thought of as representing {\it physical time}, while the three real
coordinates
\be{spacecoord}
	{\bf X}_a = 2 \varepsilon_{abc} p_b q_c
\ee
constitute the radius vector ${\bf X}=\{X_a\}$ of the Euclidean 
{\it physical 3-space}.

Under 3-rotations (\ref{aut}) by a {\it real-valued\/} angle, the coordinate
$T$ is invariant while the coordinates ${\bf X}$ transform as an ordinary
3-vector. Therefore, we can always use these rotations to fix the plane
formed by a pair of vectors $\{ \bf p, \bf q \}$ so that one has, say, 
$z_3 = 0$, $X_1 = X_2 = 0$, whereas the real motion takes place in the  
direction $X_3$ orthogonal to the plane. One can then easily check  
that rotations in the
$\{z_1, z_2\}$ plane by an {\it imaginary\/} angle $i\psi$ give rise to
the corresponding {\it Lorentz boosts\/} for the quantities $\{T, X_3\}$,
yet by {\it a double (hyperbolic) angle\/} $2\psi$. Thus, Lorentz 
symmetry here manifests itself in a rather nontrivial way. 

The main invariant (\ref{MMinvar}) of the $SO(3,\mathbb{C})$ transformations can be
thus identified with the metric interval of Minkowski space. However, being
induced by a mapping from a 3D complex biquaternionic subspace, it is always
positive-definite, i.e. corresponds to timelike or null separations of the
corresponding points. Thus the existence of a primary complex space gives
rize not only to the real (1+3) pseudo-Euclidean geometry, but also to the
latter's {\it causal structure} which, in SR theory, has to be postulated
additionally.

If one now applies the above-constructed mapping to {\it increments\/}
of the corresponding coordinates, then, for an infinitesimal spacial
displacement $\delta {\bf X}$, one obtains
\be{displace}
	\delta {\bf X} = 2  \delta {\bf p} \times \delta {\bf q},
\ee
whereas for corresponding time interval  $\delta T$
one gets the positive-definite expression
\be{increm}
	\delta T = |\delta{\bf p} |^2 + |\delta {\bf q}|^2,
\ee
so that the resulting time appears to be {\it irreversible}. On the other
hand, the expression (\ref{increm}) does not form a full differential
(i.e., one here deals with an effective {\it non-holonomity} of the
resulting space-time). As a result, any {\it closed} path ${\bf z}={\bf
z}(\lambda)=\{{\bf p}(\lambda),{\bf q}(\lambda)\}, \lambda \in \C $ of a
point particle in the primary 3D complex space corresponds to a nonzero
positive value of the physical time passed
\be{interval}
   \Delta T = \oint \delta T ({\bf p}(\lambda),{\bf q}(\lambda)) d\lambda.
\ee
This effect could be closely related to that of time delay in SR. As to the
analogous effect of {\it spatial} non-holonomity, again due to bilinearity
of the corresponding expression (\ref{displace}), it has no analogues in SR
and deserves a special consideration. Generally, however, we expect that all
the familiar effects specific of SR and based completely on the Lorentz
invariance property of the Minkowski interval will be preserved in the
suggested algebro-geometric scheme.

Let us now discuss the problem of restoration of the ``hidden'' complex space
structure from that of Minkowski space-time (\ref{displace}), (\ref{increm}).
It is noteworthy that the {\it 3D complex null cone} subspace
\be{3Dcone}
	\delta \sigma =  (\delta z_1)^2 + (\delta z_2)^2 + (\delta z_3)^2 = 0
\ee
is mapped into the real light cone of Minkowski space-time
\be{Mincone}
	\delta S^2 = (\delta T)^2 - | \delta {\bf X} |^2 = 0.
\ee
Taking in account that in this case eq. (\ref{3Dcone}) leads to
the constraints
\be{constr}
	| \delta {\bf p} | = | \delta {\bf q} |, ~~
			\delta {\bf p} \cdot \delta {\bf q} = 0,
\ee
one finds that, in order to restore the complex structure (i.e. the vectors
$\delta {\bf p}, \delta {\bf q}$) from the real displacement $\delta {\bf
X}$ (in this case $\delta T = | \delta {\bf X} |$), it is necessary to set
the value of one additional angle (the {\it phase\/}). Indeed, the direction
of $\delta {\bf X}$ fixes the orthogonal plane containing both vectors
which, as a result of (\ref{constr}), are equal in modulus and mutually
orthogonal. Thus they are fixed by $\delta {\bf X}$ {\it up to a common
1-parameter rotation in this plane}.

To pass to the general case, let us recall that the main invariant $\sigma$
given by Eq. (\ref{3Dmetrics}) is complex-valued, and, apart from the 
Minkowski interval, there exists another quantity, the phase $\alpha$ 
defined by the relation
\be{phase}
\tan \alpha = \frac{2 \delta {\bf p} \cdot \delta {\bf q}}
{| \delta {\bf p}|^2 - | \delta {\bf q}|^2};
\ee
{\it it also remains invariant under the whole 6-parameter group of Lorentz
transformations}. For this reason, the ``hidden'' geometric phase $\alpha$
accompaning any displacement in Minkowski space-time is of fundamental
importance and may probably have a direct relationship to the quantum
interference phenomena. Note that, for a lightlike motion represented by
(\ref{3Dcone}), (\ref{Mincone}), the phase $\alpha$ is indefinite since both
$\Re(\sigma)$ and $\Im(\sigma)$ turn to zero.

Let us now return to the problem of restoring the primary complex geometry
in the case when, apart from $\delta T$ and $\delta {\bf X}$, the value of
the phase $\alpha$ is given. Without going into details of calculations, we
only announce the answer: the vectors $\delta{\bf p}$ and $\delta{\bf q}$ of
a complex displacement are fixed by these data, again up to a common
rotation of this pair in the plane orthogonal to the $\delta {\bf X}$
direction. In particular, for the angle $\theta$ between these vectors one
gets
\be{angle}
	\cos^2 \theta = \frac{1-v^2}{1+v^2 \cot^2 \alpha},
\ee
$v\le 1$ being the 3D velocity of motion in the units of fundamental light
velocity $c=1$. For a point at rest $v=0$, in particular, the two vectors
${\bf p, q}$ are either parallel ($\theta=0$) or antiparallel ($\theta=\pi$),
and this duality might be related to the quntum spin properties of particles.

\section{``Hidden'' algebraic dynamics in complex-quaternionic space and
its image in Minkowski space-time}

In biquaternionic ($\B$) algebrodynamics, any matrix component of the
fundamental $\B$-field, due to the generalized Cauchy-Riemann conditions,
satisfies the {\it complex eikonal equation\/} (CEE) \cite{AD,GR95}. On the
other hand, any solution of the CEE gives rize to null congruences of
``rays'', one of which is {\it shear-free\/} \cite {Eik}. An interesting
class of null shear-free congruences (SFC) is generated by a ``virtual''
point-like charge ``moving'' along an arbitrary world line $Z = Z(\kappa)$,
$\kappa \in \C $ in 4D complex space \cite{NewmanNew,Burin}.

In this case, a fundamental SFC is formed by a bundle of null rectilinear
complex ``rays'' starting/ending on the charge. The projective twistor field
of the congruence $\{\xi,\tau\}$ at a point $Z\in \C \bf M$ is determined by
the condition
\be{twist}
	(Z-{\wh Z}(\kappa))\xi = 0, \quad\ \Longleftrightarrow \quad\
		\tau = Z\xi = {\wh Z}(\kappa)\xi.
\ee
The field preserves its value along any of the rays, depends on the radial
direction in the vicinity of a charge and is indefinite at the very point
of its location. Thus the world line of the charge is just the {\it focal
line\/} of the SFC.

The position of a charge which ``affects'' the point $Z$ is fixed, on account
of (\ref{twist}), by the {\it complex null cone\/} (CNC) equation
\be{ccone}
	\det | Z - {\wh Z}(\kappa) | = 0,
\ee
which determines the value of the corresponding parameter $\kappa$. Note
that the field $\kappa(Z)$ identically satisfies the CEE.

However, contrary to the case of real Minkowski space-time $\bf M$, in the
complex space $\C \bf M$ the CNC equation (\ref{ccone}) has, generically,
{\it a great number of roots\/} $\{\kappa_n\}$ and determines {\it an
ensemble of identical point ``particles'' ${\wh Z}_n$ which ``affect''
$Z$} by the corresponding twistor field. In our paper \cite{PIRT} these were
called {\it duplicons}.

In particular, one can take a point belonging to the world line itself, i.e., 
$Z={\wh Z}(\pi)$, and consider it as an {\it elementary observer}. From the
CNC equation (\ref{ccone}) one then obtains a set of parameters
$\kappa = \kappa_n(\pi)$ and, accordingly, an ensemble of duplicons ${\wh
Z_n}(\pi)$. Note that in a real $\bf M$ this construction is impossible since
one has there the only solution of the ``light cone'' equation
$\kappa=\pi$ (i.e. the ``retardation time'' turns to zero).

Thus, at any point of its world line ${\wh Z}(\pi)$, an elementary observer
deals essentially with a (6D real\/) subspace of a local CNC --- with {\it
observable} space-time \cite{PIRT}. In the {\it relative\/} coordinates
${\overline Z}(\pi) = {\wh Z}(\pi)- {\wh Z}(\kappa(\pi))$, the local CNC
equation reads
\be{locCNC}
	({\overline z}_1)^2 + ({\overline z}_2)^2
		+ ({\overline z}_3)^2 = ({\overline z}_0)^2
\ee
and determines the observable space {\it at a fixed instant of the (complex)
proper time of the observer $z_0(\kappa)$}. Note that the whole
``matter-like'' distribution of duplicons at this $z_0$ with
respect to the observer belongs to its local CNC space and, as we shall see
now, corresponds to all different intervals of coordinate time ${\overline
T}$ {\it at its past}.

Indeed, let us now reduce the CNC (\ref{locCNC}) to real physical space-time
making use of the mapping constructed in section 2. Taking the modulus of both
sides of (\ref{locCNC}), we get on account of
(\ref{3Dmetrics}), (\ref{Minvar}) and (\ref{MMinvar}):
\be{modul}
 {\overline T}^2 - | {\bf \overline X} |^2 = S^2 \equiv |{\overline z}_0|^2.
\ee
Thus we see that {\it the real Minkowski interval is nothing but the modulus
of the complex proper time of an elementary observer}. As to the other,
{\it phase invariant\/} (\ref{phase}), it obviously corresponds to the phase
of the complex proper time ${\overline z}_0$.

It has been shown in \cite{Levich,PIRT} that the {\it absolute} complex
coordinate $z_0$ can be considered globally as the {\it evolutional 
parameter\/} along retilinear complex rays of the null SFC which preserves
both the value of the twistor field of the congruence and {\it the caustic
structure}. In other words, the caustics propagate along {\it some\/} of the
complex rays connecting the observer with its ``images'', duplicons. On the
other hand, the coordinate $z_0$ parametrizes the location of an elementary
observer on its world line and is responsible for all alterations of the
``picture of the World'' perceived by the observer through reception of
signals-caustics from the duplicons and through permanent observation of 
the temporal dynamics of the latters correlated by the primary twistor field.

One of the peculiar features of complex time dynamics is the absence of the 
{\it event ordering} which, however, can be restored by introduction of an 
extra structure,
the {\it evolutionary curve\/} \cite{Levich,PIRT}. Along with variations of
the $\alpha$-phase invariant, this can open a way to explanation of the 
quantum uncertainty and the interference phenomena.

\section
   {Hamilton's dream on the quaternionic structure of the Universe: was it true?}

To conclude, in this paper we have proposed to consider the observable
Minkowski space-time as a ``shadow space'' of a ``hidden'' fundamental
geometry which has a purely abstract, ``numerical'' origin and is fully
encoded in the structure of the biquaternion albebra. Just in the spirit of
Hamilton, the fourth (scalar) coordinate in $\B$ is closely related to time,
but this is {\it proper time}, and this time is {\it complex-valued}. On the
other hand, Minkowski space itself, {\it together with its internal causal
structure}, is encoded in the main 3D complex subspace of $\B$ and is {\it
bilinearily} determined by the complex coordinates of the latter, through 
the structure of the $\B$-automorphism group $SO(3,\C )$. 
Moreover, another {\it phase
invariant} of the Lorentz transformations arises naturally in the approach.
It seems that we have met with quite a new ``hidden'' geometry of nature
which by itself predetermines the kinematics and interactions of
particle-like formations, their spin properties and quantum uncertainty in
general. We hope that these expectations are really well grounded.

\bigskip

\noindent
\small{{\bf Note added in pursuit.} 
Concept of the complex space-time has been proposed repeatedly in the 
frameworks of quantum mechanics (see, e.g.,~\cite{Yang, Shemizadeh}), field 
theory~\cite{Lanczos,Kaiser,Hooft} and General Relativity
(see numerous papers of R. Penrose, E.T. Newman, A. Trautman et al.).
In an algebraic approach which makes use of complex quaternions the adoption 
of this concept seems inevitable. However, up to now the {\it geometrophysical}  
sense of the additional ``imaginary'' dimensions, as well as the exceptional 
position of the Minkowski space-time in the structure of  
complexified space have not been understood at all. We hope that this 
very article will give impetus to reconsideration and intensive development 
of various approaches dealing with complexification of space-time, of the 
algebrodynamics in the first turn.}


\small
\begin{thebibliography}{99}

\bibitem{Newman}
E.T. Newman, {\it J. Math. Phys.\/} {\bf 14}, 102 (1973).

\bibitem{Lind}
R. Lind, E.T. Newman, {\it J. Math. Phys.\/} {\bf 15}, 1103 (1974).

\bibitem{Newman2}
E.T.Newman, {\it Class. Quantum Grav.}\/ {\bf 21}, 3197 (2004); gr-qc/0402056. 

\bibitem{Carter}
B. Carter, {\it Phys. Rev.\/} {\bf 174}, 1559 (1968).

\bibitem{Rob}
I. Robinson, {\it J. Math. Phys.\/} {\bf 2}, 290 (1961).

\bibitem{Tod}
K.P. Tod, {\it J. Math. Phys.\/} {\bf 23}, 1147 (1983).

\bibitem{Newman3}
E.T. Newman, {\it Phys. Rev.\/} {\bf D 65}, 104005 (2000); gr-qc/0201055.

\bibitem{AD}
V.V. Kassandrov, `Algebraic Structure of Space-Time and Algebrodynamics',
	RUDN Press, Moscow, 1992 (in Russian).

\bibitem{GR95}
V.V. Kassandrov, \GC {3} 216 (1995); gr-qc/0007027.

\bibitem{Vestnik}
V.V. Kassandrov, {\it Vestnik RUDN, Fizika} {\bf 8 (1)}, 34 (2000).

\bibitem{Jos}
V.V. Kassandrov and J.A. Rizcalla, gr-qc/0012109.

\bibitem{Sing}
V.V. Kassandrov, in: ``Has the Last Word been Said on Classical
Electrodynamics?'', eds. A. Chubykalo, V. Onoochin, A. Espinoza and
R. Smirnov-Rueda, Rinton press, 2004, p. 42; physics/0308045.

\bibitem{Acta}
V.V. Kassandrov, {\it Acta Applic. Math.\/}, {\bf 50}, 197 (1998).

\bibitem{Pavlov}
V.V. Kassandrov, {\it Hypercomplex Numbers in Geom. Phys.\/}{\bf 1(1)}, 89,
	(2004); hep-th/0312278.

\bibitem{GRG}
V.V. Kassandrov and V.N. Trishin, {\it Gen. Rel. Grav.} {\bf 36}, 1603 (2004); 
gr-qc/0401120.

\bibitem{VINITI}
V.V. Kassandrov, in: {\it VINITI Ac. Sci. USSR}, No.152-80 DEP, 1980, 11 p.
	(in Russian).

\bibitem{Tallinn}
V.V. Kassandrov, in: ``Quasigroups and Nonassociative Algebras in Physics'',
eds. J. L\~ohmus and P. Kuusk, Inst. Phys. Estonia Press, 1990, p. 202.

\bibitem{Protvino}
V.V. Kassandrov and J.A. Rizcalla, in: ``Geometrical and Topological Ideas in
Modern Physics'', ed. V.A. Petrov, IHEP Press, Protvino, 2002, p. 199.

\bibitem{Eik}
V.V. Kassandrov,  \GC {8} Suppl. II, 57 (2002); math-ph/0311006.

\bibitem{Efremov1}
A.P. Yefremov, \GC {2} 77 (1996).

\bibitem{Efremov2}
A.P. Yefremov, \GC {2} 335 (2001).

\bibitem{Urus}
I.A. Urusowskii, {\it Zarubezhnaya Radioelektronika} {\bf 3}, 3 (1996); 
{\bf 6}, 64 (1996) (in Russian).

\bibitem{Burin}
A.Ya. Burinskii, {\it Phys. Rev.\/} {\bf D 67}, 124024 (2003); gr-qc/0212048.

\bibitem{NewmanNew}
C.N. Kozameh and E.T. Newman, {\it Class. Quant. Grav.}, {\bf 22}, 4667
(2005); gr-qc/0504093.

\bibitem{Levich}
V.V. Kassandrov, in: ``Proceedings of the Seminar on Studing the Time
Phenomenon'', ed. Levich A.P., Moscow, 2006 (to appear, in Russian).

\bibitem{PIRT}
V.V. Kassandrov, in: ``Proc. Int. Conf. on Physical Interpretation of
Relativity Theory'', eds. M.C. Duffy, A.N. Morozov, V.O. Gladyshev and
P. Rowlands, Bauman Tech. State Univ. Press, Moscow, 2005, p. 42; gr-qc/0602064.

\bibitem{Yang}
C.-D. Yang, {\it Ann. Phys.}, {\bf 319} 399 (2005).

\bibitem{Shemizadeh}
V.E. Shemi-zadeh, quant-ph/0206195.

\bibitem{Hooft}
G.`t Hooft, S. Nobbenhuis, gr-qc/0602076.  

\bibitem{Lanczos}
C. Lanczos, in: ``Cornelius Lanczos Collected Published Papers with 
Commentaries'', eds. W.R. Davis et al., {\bf 6},  N. Carolina State Univ. 
Press, Raleigh, 1998, p. A-1; see also the papers of A. Gsponer, 
e.g. A. Gsponer, gr-qc/0405046.  

\bibitem{Kaiser}
G. Kaiser, {\it J. Phys. A}, {\bf 36} R291 (2003); math-ph/0303027.


\end{thebibliography}
\end{document}